# Nonadditive generalization of the quantum Kullback-Leibler divergence for measuring the degree of purification

Sumiyoshi Abe

*Institute of Physics*, *University of Tsukuba*, *Ibaraki 305-8571*, *Japan*

The Kullback-Leibler divergence offers an information-theoretic basis for measuring the difference between two given distributions. Its quantum analog, however, fails to play a corresponding role for comparing two density matrices, if the reference states are pure states. Here, it is shown that nonadditive (nonextensive) generalization of quantum information theory is free from such a difficulty and the associated quantity, termed the quantum $q$-divergence, can in fact be a good information-theoretic measure of the degree of state purification. The correspondence relation between the ordinary divergence and the $q$-divergence is violated for the pure reference states, in general.

Purification is of fundamental relevance to quantum error correction, which is important for quantum computation and quantum communication. Specifically, a task is to purify a state of a subsystem of a composite system decayed into a mixed state (see [1-4], for example).

In such a situation, it is essential to quantify the degree of purification, that is, to compare a mixed-state density matrix with a reference pure-state density matrix. This problem is often treated by the use of the concept of fidelity [5,6]. For two density matrices, $\rho$ and $\sigma$, it is given by

$$F[\sigma, \rho] = \left[ \text{Tr} \left( \sqrt{\sigma} \rho \sqrt{\sigma} \right)^{1/2} \right]^2, \tag{1}$$

which is also related to the Bures metric between $\rho$ and $\sigma$ as $d_B^2 = 2 - 2\sqrt{F[\sigma, \rho]}$. For a pure state, $\sigma = |\psi\rangle\langle\psi|$, the fidelity becomes $F[|\psi\rangle\langle\psi|, \rho] = \langle\psi|\rho|\psi\rangle$.

On the other hand, in classical information theory, comparison of two distributions is customarily discussed by employing the Kullback-Leibler divergence. Its quantum-mechanical counterpart is the quantum divergence of a density matrix $\rho$ with respect to a reference density matrix $\sigma$, which is given by [7]

$$K[\rho\|\sigma] = \text{Tr}\left[\rho(\ln\rho - \ln\sigma)\right] \geq 0, \tag{2}$$

where the equality holds if and only if $\rho = \sigma$. However, *this quantity turns out to be inadequate for measuring degree of purification, since* $\ln \sigma$ *is a singular quantity if the reference state* $\sigma$ *is a pure state*. (More generically, $K[\rho \| \sigma]$ can be well-defined only when the support of $\sigma$ is equal or larger than that of $\rho$ [7].)

In this paper, we study a generalized information-theoretic approach to quantifying the degree of purification based on nonadditive quantum information theory, which has been initiated in [8] and applied to the problems of quantum entanglement [9-11]. In particular, we discuss nonadditive generalization of the quantum divergence, termed the quantum $q$-divergence, and explicitly show how it is superior to the one in Eq. (1).

Let us start our discussion with noting that the ordinary quantum divergence in Eq. (2) can be rewritten as follows:

$$K[\rho \| \sigma] = \frac{d}{dx} \text{Tr}\left(\rho^x \sigma^{1-x}\right)\bigg|_{x \to 1-0}. \tag{3}$$

The quantum $q$-divergence is obtained by replacing the derivative in Eq. (3) with the Jackson $q$-derivative:

$$K_q[\rho \| \sigma] = D_q \text{Tr}\left(\rho^x \sigma^{1-x}\right)\bigg|_{x \to 1-0}, \tag{4}$$

where $D_q$ denotes the Jackson differential operator defined by

$$D_q f(x) = \frac{f(qx) - f(x)}{x(q-1)},\qquad(5)$$

which satisfies the following $q$-deformed Leibniz rule:

$$D_q(f(x)g(x)) = (D_q f(x))g(x) + f(x)(D_q g(x))$$
$$+ x(q-1)(D_q f(x))(D_q g(x)).\qquad(6)$$

In the limit $q \to 1$, $D_q f$ tends to the ordinary derivative, $df/dx$. Eq. (4) is found to be

$$K_q[\rho\|\sigma] = \frac{1}{1-q}\mathrm{Tr}\left[\rho^q\left(\rho^{1-q} - \sigma^{1-q}\right)\right]$$

$$= \mathrm{Tr}\left[\rho^q\left(\ln_q \rho - \ln_q \sigma\right)\right].\qquad(7)$$

In this equation, $q$ is a positive parameter termed the entropic index, and $\ln_q x$ stands for the $q$-logarithmic function defined by

$$\ln_q x = \frac{1}{1-q}\left(x^{1-q} - 1\right),\qquad(8)$$

which converges to the ordinary logarithmic function, $\ln x$, in the limit $q \to 1$. Therefore, $K_q$ might also be expected to converge to $K$ in such a limit. (However, we shall see that this is not the case, in general.) Since $K_q$ should not be too sensitive to small eigenvalues of $\rho$ and $\sigma$, the range of the entropic index must be taken to be

$$0 < q < 1. \tag{9}$$

Two comments are in order. Firstly, the classical counterpart of Eq. (7) has been proposed independently and almost simultaneously in [12-14]. Secondly, the above construction reminds us of that of the Tsallis entropy [15] developed in [16]. This is due to the fact that $K_q$ in Eq. (6) is the relative entropy associated with the Tsallis entropy, $S_q[\rho] = (\operatorname{Tr} \rho^q - 1)/(1-q)$, analogously to the relationship between $K$ in Eq. (2) and the von Neumann entropy, $S[\rho] = -\operatorname{Tr}(\rho \ln \rho)$.

$K_q$ is nonadditive in the sense that for the factorized joint density matrices of a composite system $(A, B)$, $\rho(A, B) = \rho_1(A) \otimes \rho_2(B)$ and $\sigma(A, B) = \sigma_1(A) \otimes \sigma_2(B)$, it yields

$$K_q[\rho_1 \otimes \rho_2 \| \sigma_1 \otimes \sigma_2] = K_q[\rho_1 \| \sigma_1] + K_q[\rho_2 \| \sigma_2]$$
$$+ (q-1) K_q[\rho_1 \| \sigma_1] K_q[\rho_2 \| \sigma_2], \tag{10}$$

which essentially has its origin in the $q$-deformed Leibniz rule in Eq. (6). Thus, the value of $1-q$ indicates the degree of nonadditivity.

Let us see that $K_q[\rho\|\sigma]$ is nonnegative for any two density matrices, $\rho$ and $\sigma$. For this purpose, consider the diagonal decompositions of $\rho$ and $\sigma$ [17]:

$$\rho = \sum_a r(a)|a\rangle\langle a|, \qquad \sigma = \sum_b s(b)|b\rangle\langle b|, \qquad (11)$$

where $\{|a\rangle\}$ and $\{|b\rangle\}$ are the orthonormal complete bases, $0 \leq r(a), s(b) \leq 1$, and $\sum_a r(a) = \sum_b s(b) = 1$. A straightforward calculation shows that

$$K_q[\rho\|\sigma] = \frac{1}{1-q} \sum_{a,b} |\langle a|b\rangle|^2 r(a)\left[1 - \left(\frac{s(b)}{r(a)}\right)^{1-q}\right]. \qquad (12)$$

Making use of the inequality, $(1-x^p)/p \geq 1-x$ ($x \geq 0$, $0 < p < 1$) with the equality for $x = 1$ (Theorem 42 in [18]), we arrive at the conclusion:

$$K_q[\rho\|\sigma] \geq \sum_{a,b} |\langle a|b\rangle|^2 r(a)\left[1 - \frac{s(b)}{r(a)}\right] = 0. \qquad (13)$$

Now, a point of crucial difference between $K[\rho\|\sigma]$ and $K_q[\rho\|\sigma]$ is that, in marked contrast with $\ln\sigma$, $\ln_q \sigma$ is a well-defined quantity for a pure reference state $\sigma = |\psi\rangle\langle\psi|$. In fact, $\sigma^{1-q} = \sigma$ ($0 < q < 1$). Accordingly, Eq. (7) is seen to be

$$K_q[\rho \| |\psi\rangle\langle\psi|] = \frac{1}{1-q}\left(1 - \langle\psi|\rho^q|\psi\rangle\right). \tag{14}$$

*It is important to note that the additive limit $q \to 1$ cannot be taken in this equation any more*.

Here, we wish to consider the particular case when $\rho$ is also a pure state, $\rho = |\phi\rangle\langle\phi|$. Then, Eq. (14) further becomes

$$K_q[|\phi\rangle\langle\phi| \| |\psi\rangle\langle\psi|] = \frac{d_{FS}^2}{1-q}, \tag{15}$$

where

$$d_{FS}^2 = 1 - |\langle\phi|\psi\rangle|^2 \tag{16}$$

is the Fubini-Study metric in the projective Hilbert space, which may give the geometric interpretations to quantum uncertainty and correlation [19]. In addition, the transition probability on the right-hand side of Eq. (16) coincides with the value of the fidelity in this case.

Finally, let us examine the quantum $q$-divergence for measuring the degree of purification of the Werner state [2]. The Werner state is a state of a bipartite spin-1/2 system, i.e., two qubits, given as follows [20]:

$$\rho_W = F|\Psi^-\rangle\langle\Psi^-| + \frac{1-F}{3}\left(|\Psi^+\rangle\langle\Psi^+| + |\Phi^+\rangle\langle\Phi^+| + |\Phi^-\rangle\langle\Phi^-|\right), \quad (17)$$

where $|\Psi^\pm\rangle$ and $|\Phi^\pm\rangle$ are the Bell states: $|\Psi^\pm\rangle = 2^{-1/2}\left(|\uparrow\downarrow\rangle \pm |\downarrow\uparrow\rangle\right)$, $|\Phi^\pm\rangle = 2^{-1/2}\left(|\uparrow\uparrow\rangle \pm |\downarrow\downarrow\rangle\right)$. $F$ is the fidelity with respect to the reference state $\sigma = |\Psi^-\rangle\langle\Psi^-|$. Its allowed range is $1/4 \leq F \leq 1$, and $\rho_W$ is known to be separable if and only if $F \leq 1/2$. In a recent paper [21], it has been discussed how to experimentally prepare such a state.

Now, the quantum $q$-divergence of $\rho_W$ with respect to the reference state $\sigma = |\Psi^-\rangle\langle\Psi^-|$ is immediately calculated to be

$$K_q\left[\rho_W \| |\Psi^-\rangle\langle\Psi^-|\right] = \frac{1}{1-q}(1 - F^q) \geq 0, \quad (18)$$

where the zero value is realized when $F = 1$ or $q \to +0$. However, as already stressed, the limit $q \to 1-0$ is singular and does not commute with the limit $F \to 1-0$.

In conclusion, we have discussed a possible information-theoretic measure of the degree of state purification based on nonadditive quantum information theory. We have analyzed the properties of the quantum $q$-divergence and have found that, for pure reference states, it is superior to the ordinary quantum divergence. In particular, we have seen that the additive limit cannot be taken in such a situation, and the correspondence relation between the ordinary divergence and the $q$-divergence is violated.

The author thanks Dr. A. K. Rajagopal for discussions.